# Experimental Observation of Efficient Nonreciprocal Mode Transitions via Spatiotemporally-Modulated Acoustic Metamaterials

Zhaoxian Chen[1,4,*], Yugui Peng[2,*], Haoxiang Li[1,*], Jingjing Liu[1], Yujiang Ding[1], Bin Liang[1,†], Xuefeng Zhu[3,†], Andrea Alu[2,†], Yanqing Lu[4], Jianchun Cheng[1,†]

[1]Key Laboratory of Modern Acoustics, MOE, Institute of Acoustics, Department of Physics, Nanjing University, Nanjing 210093, People's Republic of China

[2]Photonics Initiative, Advanced Science Research Center, City University of New York, New York 10031, USA

[3]School of Physics and Innovation Institute, Huazhong University of Science and Technology, Wuhan, Hubei 430074, People's Republic of China

[4]College of Engineering and Applied Sciences, Nanjing University, Nanjing 210093, People's Republic of China

In lossless acoustic systems, mode transitions are always time-reversible, consistent with Lorentz reciprocity, giving rise to symmetric sound manipulation in space-time. To overcome this fundamental limitation and break space-time symmetry, nonreciprocal sound steering is realized by designing and experimentally implementing spatiotemporally-modulated acoustic metamaterials. Relying on no slow mechanical parts, unstable and noisy airflow or complicated piezoelectric array, our mechanism uses the coupling between an ultrathin membrane and external electromagnetic field to realize programmable, dynamic control of acoustic impedance in a motionless and noiseless manner. The fast and flexible impedance modulation at the deeply subwavelength scale enabled by our compact metamaterials provides an effective unidirectional momentum in space-time to realize irreversible transition in *k*-*ω* space between different diffraction modes. The nonreciprocal wave-steering functionality of the proposed metamaterial is elucidated by theoretically deriving the time-varying acoustic response and demonstrated both numerically and experimentally via two distinctive examples of unidirectional evanescent wave conversion and nonreciprocal blue-shift focusing. This work can be further extended into the paradigm of Bloch waves and impact other vibrant domains, such as non-Hermitian topological acoustics and parity-time-symmetric acoustics.

It is of fundamental interest and practical significance in acoustics to mold the acoustic fields to form the desired wavefront, but acoustic propagation in Hermitian systems with time reversal symmetry is always reversible under the constraint of Lorentz reciprocity.[1] Due to the lack of a spin degree of freedom, magneto-acoustic effects are usually too weak to support nonreciprocal acoustic propagation as in optics.[2] Attempts to realize acoustic nonreciprocity trace back to the so-called acoustic diodes based on nonlinearities,[3-5] but they can hardly be used to realize nonreciprocal sound steering owing to the material instability, distortions, low harmonic efficiency and output phase chaos. The recent introduction of parity-time symmetry [6] for optics [7-8] and acoustics [9-12] has enabled anisotropic transmission resonances and unidirectional wave diffraction, but in these systems reciprocity cannot be broken.[13] In the past decades, progress in condensed matter physics has extended the paradigm of topological insulators from electronics [14] to photonics,[15] mechanics [16] and acoustics.[17] However, the bosonic nature of sound makes it virtually impossible to suppress reflection for reciprocal transitions between paired interface modes carrying opposite pseudospins or valleys owing to unavoidable mode coupling.[18-20] The recently-emerged acoustic Chern insulators [21-23] or Floquet insulators [24-26] give rise to topologically robust one-way sound transport induced by time-reversal symmetry breaking, but with the requirement of carefully controlled piezo-electric arrays, mechanical rotation or vibration controlled by motors and unstable air flow.[27-28] To our knowledge, nonreciprocal air-borne sound steering still remains elusive due to the challenges in the experimental realization of fast, efficient and precisely tailored space-time modulation required by the production of irreversible transition in *k-ω* space between different diffraction modes.

In this article, we overcome these limitations in terms of speed, efficiency and precision of spatiotemporal modulation in acoustics [29-31] and experimentally demonstrate irreversible mode transition and nonreciprocal steering of air-borne sound waves. With no dependence on slowly-moving mechanical components,[32-33] individually controlled piezoelectric elements in large arrays [26] or flows in bulky structures,[28] our mechanism modulates spatiotemporally the acoustic response of a vanishingly-thin resonating membrane coupled to an external electromagnetic field. As a convenient implementation, the metamaterial unit cell is designed using a thin film transducer that physically couples the electromagnetic field and the acoustic impedance, which is driven by an auxiliary lab-made circuit with programmable electronic signals. We theoretically derive the time-varying effective impedance of the proposed metamaterial unit cell and verify its acoustic response via numerical simulations and experimental measurements. Under a precisely designed periodic spatiotemporal modulation realized by adjusting the electro-acoustic coupling strength with the auxiliary circuit, the propagating acoustic mode is shown to transition unidirectionally in the *k-ω* band diagram with broken space-time reversal symmetry, as schematically shown in **Figure 1**. This enables various nonreciprocal sound steering functions such as unidirectional evanescent wave conversion and nonreciprocal blue-shift focusing. Our study opens the door to the use of spatiotemporally modulated metamaterials to produce large manipulation of sound in space-time.

Figure 1a schematically illustrates the design of time-varying metamaterials playing a pivotal role in the realization of ultrafast and efficient spatiotemporal modulation and large nonreciprocity in sound steering. Inherently, each metamaterial unit cell is designed as an ultracompact membrane, much smaller than the wavelength in all three dimensions, which is coupled to an external electromagnetic field with fast-varying and adjustable coupling strength. The capability of the membrane to interact with the incident acoustic wave is partially dependent on the coupling strength, enabling an ultrafast impedance modulation with no need for mechanical rotation or air flow. For simplicity and without loss of generality, in the current study this model is implemented by assembling a commercial loudspeaker, consisting of a deep-subwavelength thin film coupled to a coil controlled by a magnetic field, with an auxiliary programmable circuit. In contrast to the conventional operation of a loudspeaker to radiate sound through a vibrating film driven by the Lorentz force acting on the coil, here the coil serves as an additional load on the film that can be controlled by an electric relay in the circuit. When the relay switches ON, the acoustic impedance changes abruptly from $Z_i$ to $Z_i + \Delta Z$ with $Z_i$ and $\Delta Z$ being the intrinsic acoustic impedance of the thin film and the additional impedance resulting from the electro-acoustic coupling, respectively. For a 1D array of these metamaterial unit cells aligned along the $x$ direction as shown in Figure 1a, a periodic modulation of acoustic impedance can be easily implemented in space-time, with periods $P_m$ and $T_m$ respectively as shown in Figure 1b, by programming the circuit to feed the relays with electric square-wave signals with ordered phase sequences. Applying a Fourier expansion, the surface impedance is expressed with a series of sinusoidal components carrying different effective momenta, as

$$Z(x,t) = Z_0 + \sum_n Z_n \cos(n\omega_m t - nk_m x), \qquad (1)$$

where $n$ equals to odd integers, $\omega_m = 2\pi/T_m$ is the modulating frequency, $k_m = 2\pi/P_m$ is the modulating spatial wave vector, $Z_0$ is the static acoustic impedance and $Z_n$ is the modulation strength of the $n$-th order sinusoidal component (see Note 1 in Supporting Information for detailed deduction). Considering that the first order component dominates the response, as shown in Figure 1c, we mainly focus on the first-order frequency transition. When a plane wave mode with frequency $\omega_i$ and wave vector $k_i$ impinges on the metamaterial surface, the transmitted plane wave modes will contain both blue-shift and red-shift parts due to the modulation,[34-40] which is expressed, up to the first order approximation, as

$$p(\mathbf{r},t) = p_0 \exp(i\omega_i t - i\mathbf{k}_i \cdot \mathbf{r}) + p_1 \exp[i(\omega_i + \omega_m)t - i\mathbf{k}_i \cdot \mathbf{r} - ik_m x] + \\ p_{-1} \exp[i(\omega_i - \omega_m)t - i\mathbf{k}_i \cdot \mathbf{r} + ik_m x], \qquad (2)$$

where $p_0$, $p_1$, and $p_{-1}$ are the pressure amplitudes of the 0th, 1st, and −1st ordered modes. Here, the effective momentum induced by spatiotemporal modulation breaks time-reversal symmetry and it is crucial for nonreciprocal mode transition. For example, in Figure 1d-e, when the acoustic plane wave mode at $\omega_i$ and $k_x = 0$ impinges normally on the spatiotemporally modulating metamaterial surface, the blue-shift mode at $\omega_i + \omega_m$ acquires an additional momentum $k_m$ along the $x$ direction and leaves the metamaterial at a refraction angle $\theta_1$. However, when the plane wave mode at $\omega_i + \omega_m$ propagates backwards in the reversed direction, the mode after red shift carries the frequency of $\omega_i$, and its momentum will decrease by $k_m$, transforming it into an evanescent mode. Since the transmitted mode of $\omega_i$ cannot leave the surface vertically along the route that it initially enters, a nonreciprocal mode transition occurs.

Due to the deeply subwavelength nature of the designed metamaterial unit cell, the thin film vibrates like a piston, which facilitates the analysis of the mode transition in terms of the time-modulation process.[41] Modeling the acoustic impedance in terms of lumped parameters,

$$Z_i = \left[\delta_m + i\left(M_m\omega - \frac{K_m}{\omega}\right)\right]\bigg/S_d , \tag{3a}$$

$$\Delta Z = (BL)^2 / Z_e S_d , \tag{3b}$$

where $\delta_m$, $M_m$, $K_m$ are the mechanical damping coefficient, mass and stiffness of the thin film, respectively, $S_d = \pi d^2/4$ is the radiation area determined by the diameter $d$, $Z_e$ is its electric impedance, and the force factor $BL$ can be used for ultrafast modulation of the electro-acoustic coupling strength. All these parameters can be easily measured or calculated by the linear parameter measuring system of Klippel [42] and are included in Methods. In experiments, as schematically shown in Figure 2a-b, the sample has a diameter of $d = 2\,\text{cm}$, which is integrated in a 3D-printed holder and fixed in the impedance tube with an inner diameter of 5 cm. The relay in the transducer's circuit could switch between on and off and is controlled by the single-chip microcomputer. Therefore the transducer's impedance is modulated at will. The sample has a resonance frequency at around 970 Hz, where the acoustic impedances are predicted from Equation 3 to be $Z_i = 95\,\text{N}\cdot\text{s/m}^3$ and $Z_i + \Delta Z = 131\,\text{N}\cdot\text{s/m}^3$ respectively before and after the relay switches on. The transmission and reflection spectra of the sample are simulated by COMSOL Multiphysics. At 970 Hz, as shown in Figure 2c, the transmission peak decreases from 0.53 to 0.45, while the reflection spectrum increases from 0.47 to 0.55 after the relay switches on. The peak frequency nearly does not shift due to the small inductance of the sample. We also experimentally measured the transmission and reflection based on the four-microphone method.[43] Figure 2d shows the measurement result with the transmission peak down from 0.58 to 0.44 and reflection up from 0.47 to 0.61, in a fairly good agreement with simulations. The results above verify that the proposed metamaterial unit-cell impedance can be effectively modulated by the driving circuit as expected.

To achieve an effective mode transition in space-time, the sound frequency is set to be 970 Hz, where the relay operates under an electric square-wave signal of 100 Hz. As shown in Figure 2e, the measured transmission spectrum has both blue-shift and red-shift modes at 970 ± 100$n$ Hz. With a square-wave modulation, shown in the inset, the odd-order modes are much stronger than the even-order ones (more details in Supplementary Note 2). The finite-difference time-domain simulations agree well with the experimental measurements, where the measured

SPLs are slightly lower than the simulation results. This may result from unavoidable relaxation effects in electro-acoustic coupling.

Next, we demonstrate nonreciprocal wave-steering functionalities enabled by our proposed time-varying acoustic metamaterials both numerically and experimentally, via two distinct examples of unidirectional evanescent wave conversion and nonreciprocal blue-shift focusing. The sample is aligned along the *x*-direction, with the initial phase of impedance modulation implementing a gradient $d\varphi/dx$ distribution shown in Figure 3a, which physically serves as an effective unidirectional momentum. Based on the conservation of momentum, the generalized Snell's law for the 1st ordered blue-shift mode follows

$$k_i \sin(\theta_i) + d\varphi/dx = k_1 \sin(\theta_1), \tag{4}$$

where $k_i$ and $\theta_i$ are the wave vector and incident angle of the incident waves at the frequency of $\omega_i$, $k_1$ and $\theta_1$ are the wave vector and refracting angle of the 1st ordered mode at $\omega_i + \omega_m$. We calculate the relation between $\theta_i$ and $\theta_1$ and plot some typical results as the green curve in Figure 3b where $\omega_i/2\pi = 1000$ Hz, $\omega_m/2\pi = 100$ Hz and $d\varphi/dx = 2\pi/60$ cm$^{-1}$. However, when we reverse the outgoing waves back as incident waves at $\omega_i/2\pi = 1100$ Hz, the mode can transit back to red-shifted modes at $(\omega_i - \omega_m)/2\pi = 1000$ Hz, losing the momentum to obey the conservation of momentum

$$k_i \sin(\theta_i) - d\varphi/dx = k_{-1} \sin(\theta_{-1}), \tag{5}$$

where $k_{-1}$ and $\theta_{-1}$ are the wave vector and refracting angle of the −1st ordered mode. The relation between $\theta_i$ and $\theta_{-1}$ is numerically calculated and shown as the red curve in Figure 3c. We also calculate the −1st ordered mode of 900 Hz as the yellow curve in Figure 3b and the 1st ordered mode of 1200 Hz as the blue curve in Figure 3c. From Figure 3b-c, we see that a nonreciprocal mode transition is obtained for wide-angle incidence, $\theta_i = -1° \sim 32°$. For example, as marked by the black dots in Figure 3b-c, for the forward propagating mode at $\theta_i = 0°$ and $\omega_i/2\pi = 1000$ Hz, the refraction angle and frequency of blue-shift outgoing mode are $\theta_1 = 32°$ and $(\omega_i + \omega_m)/2\pi = 1100$ Hz. By reversing the outgoing mode into backward at $\theta_i = -32°$ and $\omega_i/2\pi = 1100$ Hz, the red-shift mode will operate at $\theta_{-1} = -90°$ and $(\omega_i - \omega_m)/2\pi = 1000$ Hz, which turns into an evanescent mode along the metamaterial surface, instead of leaving the surface in the normal direction. In experiments, the plane wave is generated by 8 loudspeakers equally spaced by 10 cm, and the metamaterial is placed 30 cm away from the line source in parallel (more details in Supplementary Note 3). The measured results agree well with the simulated results as shown in Figure 3d-e and clearly display the phenomenon of unidirectional evanescent wave conversion,[44] breaking the intrinsic limitations in previous metamaterial-based wave manipulation. To provide a full view of the mode transition process, Figure 3f-g show the −1st ordered mode of 900 Hz for the forward propagation and the 1st ordered mode of 1200 Hz for the backward propagation, respectively. This nonreciprocity allows efficient radiation of the acoustic energy along the target direction, while guiding the reversed incident waves as surface waves, opening the possibility for the design of novel functional devices such as nonreciprocal phased arrays with applications in acoustic imaging and communications.

Last, we adapt the time-varying metamaterial into an acoustic meta-lens for nonreciprocal blue-shift focusing. For a plane wave mode at 1000 Hz normally impinging on the metamaterial, we design an initial phase distribution of impedance modulation that can focus the blue-shift 1st order mode at 1100 Hz into the prescribed focal point. Here the phase distribution follows [45]

$$\varphi(x) = k_1(\sqrt{x^2 + y^2} - y) , \qquad (6)$$

where the focal point is located at $y = 30$ cm and the phase distribution is shown in Figure 4a. In experiments, the phase lag between adjacent unit-cells is $\pi/8$. The blue-shift focusing effect is verified in simulations and experiments, as shown in Figure 4b, in good agreement with each other. To quantitatively investigate the focusing effect, we plot the normalized intensity distribution along the cut line at $y = 25$ cm, which is the real focal position in the measurement, as shown in Figure 4c. The full width at the half magnitude of the blue-shift focal spot is around 16 cm, less than half wavelength (or breaking the diffraction limit) of sound at 1000 Hz (more details in Supplementary Note 5). Here the real focal position is closer to the sample surface than the designed one due to the discrete phase distribution instead of a continuous one. In the reverse case, if we put a point source of 1100 Hz at the focal point of $x = 0$ cm and $y = 25$ cm, the red-shift mode at 1000 Hz will turn into divergent beams instead of the collimated beams in Figure 4d, indicating that the sound focusing is nonreciprocal. Such unconventional functionality can have far-reaching implications in important scenarios ranging from ultrasound therapy to nondestructive evaluation, which calls for high-efficiency conversion of acoustic energy but traditionally suffers from undesired reversed wave propagation back to the source.

In conclusion, we have demonstrated the possibility of nonreciprocal mode transition in time-varying acoustic metamaterials. By meticulously tailoring the spatiotemporal modulation, effective unidirectional momentum has been generated in a time-varying system and irreversible mode transition has been shown in *k-ω* space. We have successfully realized various intriguing nonreciprocal phenomena, such as unidirectional evanescent wave conversion and nonreciprocal blue-shift focusing, which are impossible in reciprocal acoustic systems. It should be mentioned that, compared with other time-varying modulation methods, such as mechanical vibrations or flow circulation, the electrically controlled switching is featured with various advantages in implementing giant and tailorable nonreciprocal transitions. First of all, the modulation frequency of our approach can in principle go up to 1000 Hz due to the fast relay response. In addition, without the additional noise from vibrating or rotating components, the proposed system becomes more robust in wave space-time manipulation. Nonetheless, the mode transition efficiency in this work can be further improved by increasing the force factor *BL* or introducing resonant meta-structures. Our proposal can be flexibly merged with 3D acoustic time-varying metamaterials, which will shed light on nonreciprocal topological acoustics, or acoustic physics with synthetic dimensions.

**Experimental Section**

*Experimental Details:* We use commercial coil-moving electroacoustic membrane transducers as the unit cells and their linear parameters, as shown in Table 1, are measured by the small signal Klippel system. The relays of SIP-1A05 are used as switches to control the impedance of the metamaterial. In the impedance tube experiment (Figure S4 in Supplementary Note 2), four 1/4-inch-diameter microphones (Brüel & Kjaer type-4961) and a multichannel analyzer (Brüel & Kjaer Pulse Type 3160) are used to extract the pressure information. In the 2D sound field steering experiment, a single-chip computer (Arduino Mega 2560) is used to provide multichannel square wave voltage signals with elaborately programmed frequency and phase lag for controlling the relays. The measurement is performed in a 2D parallel waveguide made of acrylic boards (more details in Supplementary Note 3).

*Simulation Details:* A commercial FEM software (COMSOL Multiphysics) is used for the numerical study. For obtaining the simulated transmission and reflection in Figure 2c-d, the acoustic module and frequency domain solver are used with the ultrathin transducer treated as an impedance boundary calculated according to Equation 3. The mass density and sound speed of air are set as $\rho_0 = 1.21 \text{ kg/m}^3$ and $c_0 = 343 \text{ m/s}$ respectively. For the transmitting SPL analysis, we first use transient solver and physics module to get the transmitting pressure information and then fast Fourier transformation to calculate its spectrum. All the simulated spatial distributions of acoustic field in both this paper and the Supplementary Notes are also obtained by using the frequency domain solver.

**Acknowledgments**

Z.C., Y.P. and H.L contributed equally to this work. This work was supported by the National Key R&D Program of China (Grant No. 2017YFA0303700), the National Natural Science Foundation of China (Grant Nos. 11634006, 11674119, 11690030, 11690032, 11374157 and 81127901), the National Science Foundation, Simons Foundation and Department of Defense of USA, the Innovation Special Zone of National Defense Science and Technology, High-Performance Computing Center of Collaborative Innovation Center of Advanced Microstructures and A Project Funded by the Priority Academic Program Development of Jiangsu Higher Education Institutions.**References**

[1] H. Nassar, B. Yousefzadeh, R. Fleury, M. Ruzzene, A. Alù, C. Daraio, A. N. Norris, G. Huang, M. R. Haberman, *Nat. Rev. Mater.* **2020**, 5, 667.

[2] A. Yariv, P. Yeh, *Optical waves in crystal propagation and control of laser radiation.* (Weley, **1983**).

[3] B. Liang, B. Yuan, J. C. Cheng, *Phys. Rev. Lett.* **2009**, 103, 104301.

[4] B. Liang, X. S. Guo, J. Tu, D. Zhang, J. C. Cheng, *Nat. Mater.* **2010**, 9, 989.

[5] N. Boechler, G. Theocharis, C. Daraio, *Nat. Mater.* **2011**, 10, 665.

[6] C. M. Bender, S. Boettcher, *Phys. Rev. Lett.* **1998**, 80, 5243.

[7] H. Zhao, L. Feng, *Natl. Sci. Rev.* **2018**, 5, 183.

[8] M.-A. Miri, A. Alù, *Science* **2019**, 363, eaar7709.

[9] X. Zhu, H. Ramezani, C. Shi, J. Zhu, X. Zhang, *Phys. Rev. X* **2014**, 4, 031042.

[10] T. Liu, X. Zhu, F. Chen, S. Liang, J. Zhu, *Phys. Rev. Lett.* **2018**, 120, 124502.

[11] R. Fleury, D. L. Sounas, A. Alù, *Phys. Rev. Lett.* **2014**, 113, 023903.

[12] R. Fleury, D. Sounas, A. Alù, *Nat. Commun.* **2015**, 6, 5905.

[13] L. Feng, Y.-L. Xu, W. S. Fegadolli, M.-H. Lu, J. E. B. Oliveira, V. R. Almeida, Y.-F. Chen, A. Scherer, *Nat. Mater.* **2013**, 12, 108.

[14] X.-L. Qi, S.-C. Zhang, *Rev. Mod. Phys.* **2011**, 83, 1057.

[15] L. Lu, J. D. Joannopoulos, M. Soljačić, *Nat. Photonics* **2014**, 8, 821.

[16] J. Cha, K. W. Kim, C. Daraio, *Nature* **2018**, 564, 229.

[17] G. Ma, M. Xiao, C. T. Chan, *Nat. Rev. Phys.* **2019**, 1, 281.

[18] Y. G. Peng, C. Z. Qin, D. G. Zhao, Y. X. Shen, X. Y. Xu, M. Bao, H. Jia, X. F. Zhu, *Nat.*9

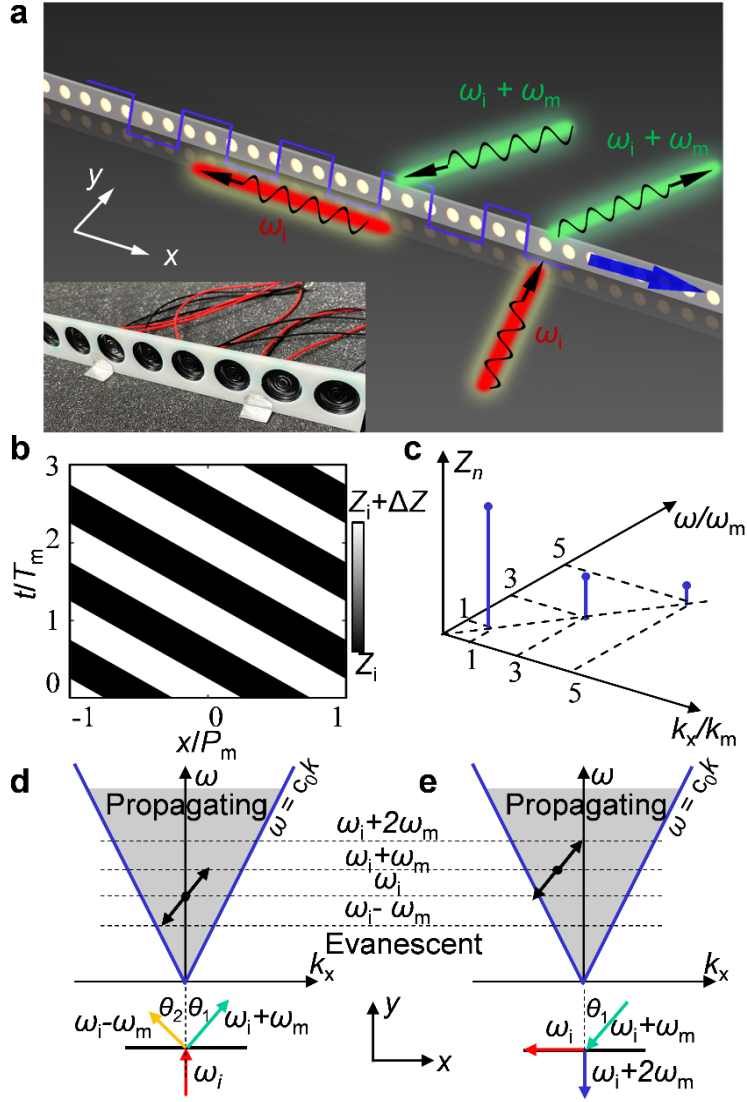

**Figure 1.** Mechanism of the spatiotemporally-modulated metamaterials. a) Schematic of nonreciprocal sound steering with the proposed space-time modulating metamaterials. Inset: a photo of the implemented 1D array of metamaterial unit cells each comprising a thin-film transducer and auxiliary electronic circuit, allowing fast and high-efficiency modulation of effective acoustic impedance by controlling voltages. b) Spatiotemporal modulation of impedance of the metamaterial in time and space, showing that by switching off and on the relay of each element, the corresponding impedance is tuned between $Z_i$ and $Z_i+\Delta Z$ with space and time periods $P_m$ and $T_m$, respectively. c) Fourier series expansion of impedance modulation in $k$-$\omega$ dimensions. d-e) Illustration of nonreciprocal mode transition for waves travelling along the d) forward direction and e) backward direction respectively. The acoustic mode at $\omega_i$ and $k_x = 0$ transits to $\omega_i + \omega_m$ with $k_x = k_m$ and $\omega_i - \omega_m$ with $k_x = -k_m$, respectively, in the forward direction. In the backward direction, the mode at $\omega_i + \omega_m$ and $k_x = -k_m$ transits back into $\omega_i$ which now is in the evanescent regime below the air line.

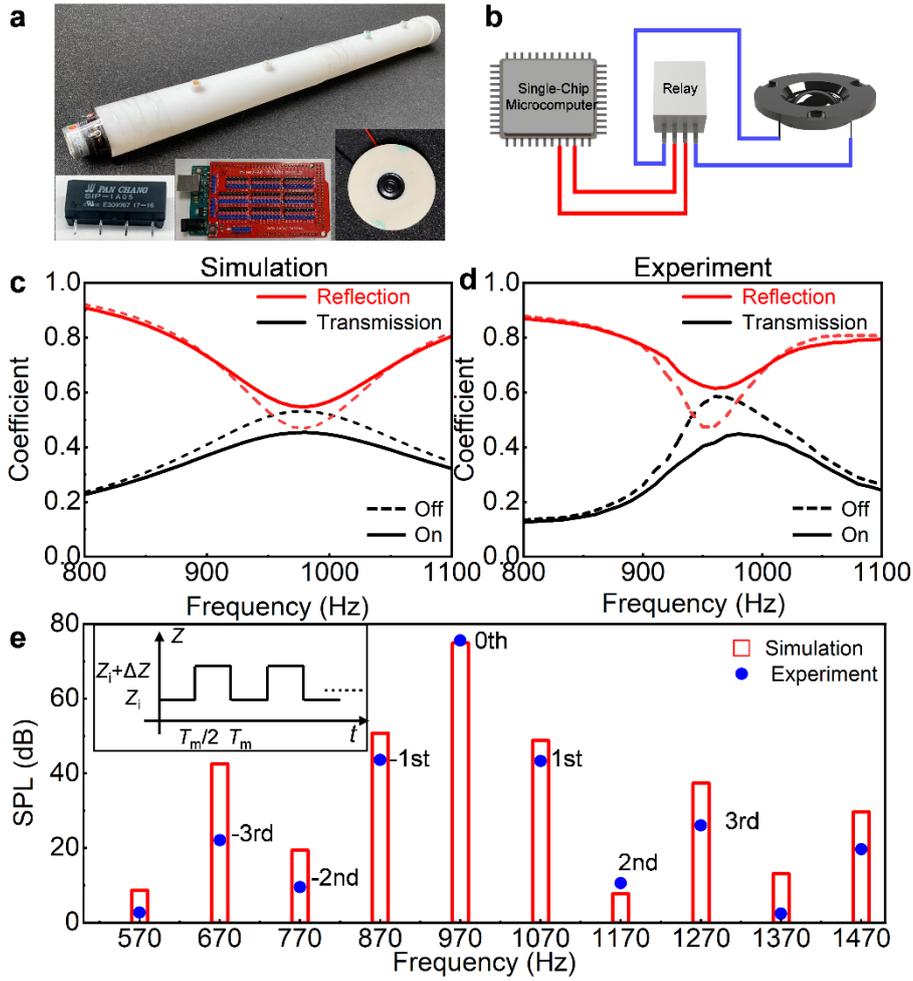

**Figure 2.** Frequency conversion study. a) Lab-designed impedance tube with inner diameter of 5 cm for measuring the sound properties of the designed time-modulated metamaterial unit cell. A loudspeaker is fixed at one end, and absorptive foam is placed at the other end to eliminate reflections. Four holes are drilled for microphones to extract the pressure and phase information. Insets from left to right are the relay, single-chip microcomputer and the fabricated sample of metamaterial unit cell, respectively. The transducer has a diameter of 2 cm and is clamped in the impedance tube together with a 3D printed holder. b) The working principle for the electrical-control impedance modulation. The outer two feet of the relay connect to the transducer's electrodes, while the inner two feet connect to the single-chip microcomputer. c) Simulated and d) measured transmission and reflection coefficients of the sample with the relay switching on and off. e) Simulated and measured transmitting sound pressure level (SPL) at the operation frequency and blue/red shift frequencies. Here the operation frequency is 970 Hz and the sample is modulated at 100 Hz by a square-wave voltage change.

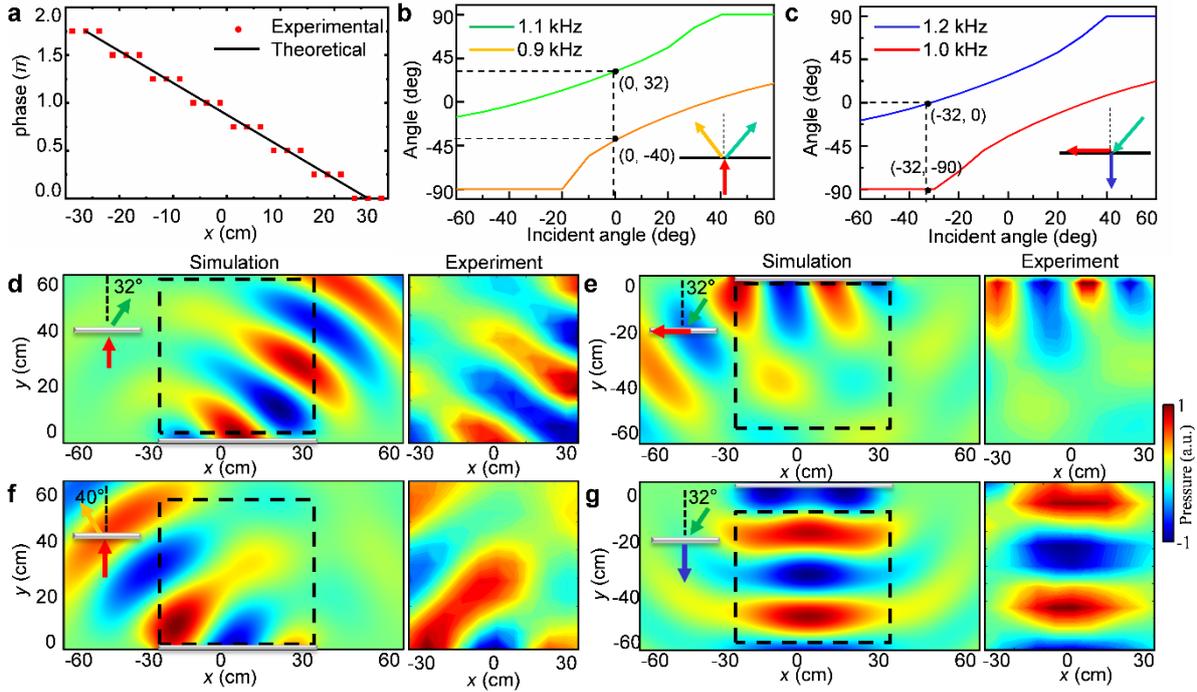

**Figure 3.** Unidirectional evanescent wave conversion. a) Initial phase distribution of the impedance modulation to generate unidirectional evanescent wave conversion. The metamaterial contains 24 time-varying unit-cells, which is 60 cm in length. The adjacent three unit-cells are set as a subgroup, and the phase lag is $\pi/4$. b) Refraction angles for outgoing acoustic wave at 1100 Hz (green curve) and 900 Hz (yellow curve) with forward incidence of 1000 Hz. c) Refraction angles for outgoing acoustic wave at 1200 Hz (blue curve) and 1000 Hz (red curve) with backward incidence of 1100 Hz. d-g) Simulated and measured pressure field distributions for the nonreciprocal mode transition, which corresponds to the four cases marked by the black dots in (b-c). The dashed boxes mark the measurement areas.

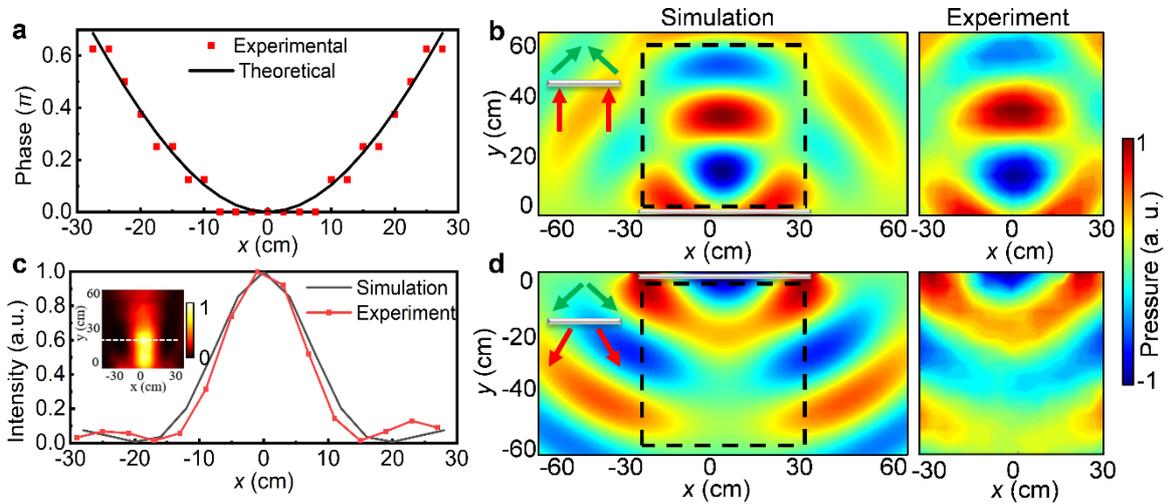

**Figure 4.** Nonreciprocal blue-shift focusing. a) Initial phase distribution of the impedance

modulation to generate nonreciprocal blue-shift focusing. The focus is designed at $y = 30$ cm behind the metamaterial and the phase lag in the sample is $\pi/8$. b) The simulated and measured focusing fields at 1100 Hz for the normal incidence of plane wave mode at 1000 Hz. c) Sound intensity distribution along the cut line at $y = 25$ cm in (b), which is the focal position. Inset figure plots the normalized intensity field. d) Simulated and measured field diverging fields at 1000 Hz for the reversed incidence in the backward direction at 1100 Hz, with a point source placed at $x = 0$ cm and $y = 25$ cm.

**Table 1.** Small signal parameters of the transducer sample

| Parameter | Notation | Value | Unit |
|---|---|---|---|
| DC resistance | $R_e$ | 8.5 | Ω |
| Coil Inductance | $L_e$ | 0.03 | mH |
| Force Factor | $BL$ | 0.3 | N/A |
| Moving Mass | $M_m$ | 0.05 | g |
| Mechanical Resistance | $\delta_m$ | 0.027 | Kg/s |
| Mechanical Stiffness | $K_m$ | 1.86 | N/mm |
| Resonating Frequency | $f_0$ | 970 | Hz |